\documentstyle[12pt]{article}

\begin{document}

\LaTeX{}\bigskip\ \bigskip\ \bigskip\ 

\begin{center}
The equation of state and specific heat of the electron gas on a one
dimensional lattice\bigskip\ \bigskip\ 

Vladan Celebonovic\medskip\ 

Institute of Physics,Pregrevica 118,11080 Zemun-Beograd,Yugoslavia\medskip\ 

vladan@phy.bg.ac.yu

vcelebonovic@sezampro.yu\bigskip\ \bigskip\ \ 

to appear in Phys.Low-Dim.Structures.,{\bf 1/2},XXX (1999) \bigskip\bigskip\
\ 
\end{center}

Abstract:Using previous results and general thermodynamical formalism,an
expression is obtained for the specific heat per particle under constant
volume of a degenerate non-relativistic electron gas on a 1D lattice.The
result is a non-linear function of the temperature,and it could have
applications in studies of quasi one-dimensional organic metals.

\newpage\ 

Introduction\medskip\ 

The purpose of this paper is to determine the equation of state and specific
heat of the electron gas on a one dimensional lattice.In experiments,the
electronic specific heat is determined indirectly, only after somehow
estimating and substracting the phonon heat capacity from the measured heat
capacity of the specimen.On the purely theoretical side,the electronic
specific heat is directly related to the exact single particle Green's
functions [1a].The calculations,to be reported in the following section,are
motivated by recent theoretical and experimental results on transport
processes in quasi one-dimensional (Q1D) organic metals [1],[2]. A
preliminary version of the calculation has already been published [3].

Q1D organic metals,later named the Bechgaard salts,were discovered in 1980
[4],[5].It was experimentally shown soon afterwards that their transport
properties are widely different from those of ordinary metals ( for example
[6] and references given there).For a general review of the field of organic
conductors see [7].\medskip\ 

The calculations\medskip\ 

It will be assumed in the calculations that the number of particles N in the
system is a variable.The starting point is a well known thermodynamical
relation [8]

\begin{equation}
\label{(1)}dG=-SdT+VdP+\mu dN 
\end{equation}

where $G=\mu N$ and all the symbols have their standard meanings. Inserting
the definition of G into eq.(1), it follows that

\begin{equation}
\label{(2)}-SdT+VdP+\mu dN=\mu dN+Nd\mu 
\end{equation}

Differentiating eq.(2) with respect to $T$ gives

\begin{equation}
\label{(3)}V(1+\frac 1V\frac{\partial V}{\partial T})\frac{\partial P}{%
\partial T}=S+\frac{\partial S}{\partial T}+\frac{\partial N}{\partial T} 
\frac{\partial \mu }{\partial T}+N\frac{\partial \mu }{\partial T} 
\end{equation}

In the last equation we have used $dP=\frac{\partial P}{\partial T}dT$ and
similar relations for other variables in eq.(2) .\\

\newpage\ 

Equation (3) is expressed in terms of the parameters of the system in
bulk.Changing the variables to the number density $n$ $(N=nV)$ and the
entropy per particle $s$ ( $S=nsV)$ after some algebra one arrives at

\begin{equation}
\label{(4)}V(1+\frac 1V\frac{\partial V}{\partial T})\frac{\partial P}{%
\partial T}=ns(V+\frac{\partial V}{\partial T})+V\frac{\partial n}{\partial T%
}(s+\frac{\partial \mu }{\partial T})+nV\frac \partial {\partial T}(s+\mu
)+n \frac{\partial \mu }{\partial T}\frac{\partial V}{\partial T} 
\end{equation}

\medskip\ 

We have thus obtained the EOS of any material in differential form.In
applications to Q1D organic metals,it can be considerably simplified.
Experiments on these materials are usually performed under constant 
volume [7]. This implies that all terms in eq.(4) containing derivatives 
of the volume can be disregarded.Accordingly,the final form of the EOS 
of Q1D organic metals is

\begin{equation}
\label{(5)}\frac{\partial P}{\partial T}=(n+\frac{\partial n}{\partial T}%
)(s+ \frac{\partial \mu }{\partial T})+n\frac{\partial s}{\partial T} 
\end{equation}

The specific heat per particle is given by [8]

\begin{equation}
\label{(6)}c_V=\frac Tn(\frac{\partial ^2P}{\partial T^2})_V-T(\frac{%
\partial ^2\mu }{\partial T^2})_V 
\end{equation}

Differentiating eq.(5),and inserting the result into eq.(6),gives the
following final expression for the specific heat per particle

\begin{equation}
\label{(7)}c_V=\frac Tn(s+\frac{\partial \mu }{\partial T})(\frac{\partial n 
}{\partial T}+\frac{\partial ^2n}{\partial T^2})+\frac Tn\frac{\partial n}{%
\partial T}(2\frac{\partial s}{\partial T}+\frac{\partial ^2\mu }{\partial
T^2})+T(\frac{\partial s}{\partial T}+\frac{\partial ^2s}{\partial T^2}) 
\end{equation}

Discussion

Equation (7) is considerably more complex than the preliminary result
already published in [3].This difference is due several simplifying
assumptions made in [3],which were dropped in the present paper.Basically,it
was assumed there that the entropy per particle is temperature
independent,that the total number of particles in the system is a
constant,and the term $\frac{\partial n}{\partial T}\frac{\partial \mu }{%
\partial T}$ was neglected.

\newpage\ 

The aim of the calculation reported here was to obtain an expression for the
differential form of the EOS and the specific heat ,but without all these
simplifications.The result for the EOS is given by eq.(4).It was derived
using general thermodynamical arguments,and accordingly can be applied to
any material.In the special case of Q1D organic metals,one can introduce the
constant volume condition into eq.(4) and thus obtain eqs.(5) and (7).

In application to the degenerate electron gas on a 1D lattice,one has to
introduce into eq.(7) appropriate expressions for the chemical
potential,number density and entropy per particle.It has recently been shown
that the chemical potential of the electron gas on a 1D lattice is given by
[9]

\begin{equation}
\label{(8)}\mu =\frac{(\beta t)^6(af-1)\left| t\right| }{1.1029+.1694(\beta
t)^2+.0654(\beta t)^4} 
\end{equation}

where $a$ denotes the lattice constant,t the hopping, $f$ the band filling
and $\beta $ the inverse temperature.The influence of the lattice is
contained in this expression through the presence of the lattice constant $%
a. $

The entropy per electron and the number density should also be expressed so
as to take into account the existence of the lattice.Work on this problem is
currently in progress,but as a first approximation one could use existing
results for the degenerate electron gas [10],[13] .The entropy per electron
is given by

\begin{equation}
\label{(9)}s=\frac Q{n_e}\frac \partial {\partial T}F_{3/2}(\beta \mu ) 
\end{equation}

where $F_{3/2}(\beta \mu )$ is a special case of a Fermi-Dirac integral
[10], $n_e$\ denotes the electronic number density,and $Q$ is a combination
of known constants ( such as the electron mass and Planck's constant).The
number density of a degenerate electron gas at low teemperature can be
expressed in analytical form as [10]

\begin{equation}
\label{(10)}n_e\cong AT^{15/2}[1+BT^{3/2}+....] 
\end{equation}

$A,B$ again denote combinations of known constants.Inserting expressions (8)
- (10) in eq.(7) one would get the final result for the specific heat of the
electron gas on a 1D lattice.\newpage\ 

How can eq.(7) be applied in the analysis of experimental data on Q1D
organic metals? It can be used in analyzing data on the specific heat and
thermal conductivity of these materials,such as those discussed in [1].

The specific heat of a solid can be decomposed into phonon and electronic
parts.Their relative contributions to the ''summary'' specific heat of the
specimen are functions of the parameters of the system.The most important of
them are the temperature and the density.Expression (8) for the chemical
potential was derived assuming that the Fermi liquid picture can be used for
describing electrons in Q1D organic metals.This assumption is at present a
subject of active discussion ( see,for example,[11],[12] ) .Holding to this
assumption,it then follows that eq.(7) can be applied to determining the
phonon part of the specific heat from the experimental data. On the other
hand,starting from the measured values of the specific heat,and the somehow
theoretically estimated phonon contribution,one could obtain the electronic
part. Comparing this function with the predictions of eq.(7),one could test
the applicability of the Fermi liquid picture to electrons in Q1D organic
metals.

The reasoning behind applications of eq.(7) to studies of the thermal
conductivity of Q1D organic metals is essentially the same.It can be shown
from elementary considerations [14] that the thermal conductivity is
proportional to the specific heat.Taking eq.(7) in account,this implies that
the thermal conductivity of Q1D organic metals is a nonlinear function of
the temperature,in line with recent experiments [1].Again,this could be used
to test theoretical models of the organic metals.Work aiming at details of
this comparison is currently in preparation.

Acknowledgement

This paper is dedicated to the memory of the late Heinz.J.Schulz from
Laboratoire de Physique des Solides in Orsay (France) who introduced me to
the field.

\newpage\ 

References\bigskip\ 

[1] S.Belin and K.Behnia,Phys.Rev.Lett.,{\bf 79},2125 (1997).

[1a] C.M.Varma,preprint cond-mat/9607105 v2 (1996).

[2] V.Celebonovic,Phys.Low-Dim.Struct.,{\bf 3/4}, 65 (1997)

and preprint cond-mat/9702147 .

[3] V.Celebonovic,Solid State Phenomena,{\bf 61-62},135 (1998c).

[4] K.Bechgaard,C.S.Jacobsen,K.Mortensen et al.,Solid

State Comm.,{\bf 33},1119 (1980).

[5] D.Jerome,A.Mazaud,M.Ribault and K.Bechgaard,

C.R.Acad.Sc.Paris.,{\bf B290} ,27 (1980);

D.Jerome,A.Mazaud,M.Ribault and K.Bechgaard,J.Physique

(Paris),Lettres,{\bf 41},L95 (1980).

[6] D.Jerome and F.Creuzet,in: Novel Superconductivity (ed.by

S.A.Wolf and V.Z.Kresin),p.103,Plenum Press, New York (1987).

[7] T.Ishiguro and K.Yamaji,Organic Conductors,Springer Verlag,

Berlin (1990).

[8] H.E.Stanley,Introduction to phase transitions and critical phenomena,

Clarendon Press,Oxford (1971).

[9] V.Celebonovic,Phys.Low-Dim.Struct.,{\bf 11/12},25 (1996)

and preprint cond-mat/9512016.

[10] V.Celebonovic,Publ.Astron.Obs.Belgrade,{\bf 60},16 (1998a)

and preprint astro-ph/9802279.

[11] J.Voit,Rep.Progr.Phys.,{\bf 58},975 (1995).

[12] C.Bourbonnais,preprint cond-mat/9611064 v2 (1996);

A.Schwartz,M.Dressel,G.Grunner et al.,Phys.Rev.,{\bf B58},1261 (1998);

J.Moser,M.Gabay,P.Auban-Senzier,D.Jerome,K.Bechgaard

and J.M.Fabre,Eur.Phys.J.,{\bf B1}, 39 (1998) .

[13] L.D.Landau and E.M.Lifchitz,Statisticheskaya fizika,Vol.1,

Nauka,Moscow (1976).

[14] N.W.Ashcroft and N.D.Mermin,Solid State Physics,

Holt,Rinehart and Winston,London (1976).

\end{document}